\documentstyle[11pt,aaspp4]{article}
\def\lya{Ly$\alpha$~}

\begin{document}

\title{Microlensing of Quasars By Stars In Their Damped \lya Absorbers} 
\author{Rosalba Perna and Abraham Loeb}
\medskip
\affil{Astronomy Department, Harvard University, 60 Garden St.,
Cambridge, MA 02138}

\begin{abstract}
The damped \lya absorbers (DLAs) in quasar spectra are believed to be the
progenitors of present--day disk galaxies. We examine the probability for
microlensing of background quasars by stars in their DLAs.  Microlensing by
an individual star should magnify the continuum but not the broad emission
lines of the quasars.  Consequently, the equivalent width distribution of
microlensed quasars would be distorted.  We model a representative spiral
galaxy as a closed system composed of a bulge, a disk, and a halo, and
evolve the mass fraction of stars in the disk based on the observed
metallicity of DLAs at high redshifts.  The microlensing signatures are
stronger if the halo of the galaxy is made of Massive Compact Halo Objects
(MACHOs).  In this case, the distortion imprinted by microlensing on the
equivalent width distribution of quasar emission lines can be detected with
high significance in a sample of $\sim 10$ DLAs with HI column densities
$N\ga 10^{21}~{\rm cm^{-2}}$ and absorption redshifts $z_{\rm abs}\la 1$.
About a tenth of all quasars with DLAs ($N\ga 10^{20}~{\rm cm^{-2}}$) might
show excess variability on timescales shorter than five years.  A search
for these signals would complement microlensing searches in local galaxies
and calibrate the MACHO mass fraction in galactic halos at high redshifts.
\end{abstract}

\keywords{cosmology: theory -- gravitational lensing --- quasars : 
emission lines}

\centerline{To appear in ApJ, 1997}

\section{Introduction}

Damped Ly$\alpha$ absorbers (DLAs) are thought to be the progenitors of
present--day disk galaxies (Wolfe 1995).  The abundance of heavy elements
at low ionization stages in the absorbers and the velocity field traced by
them is consistent with values expected for disk galaxies (Turnshek et al.
1989; Wolfe et al.  1993; Lu et al. 1993; Pettini et al. 1994; Lu \& Wolfe
1994).  Recent HIRES observations on the Keck telescope indicate that the
weak, low--ionization, metal absorption lines in these systems often show
the highest column--density component at the edges of the velocity profile,
as expected for absorption by a rotating gaseous disk (Wolfe 1995;
Prochaska \& Wolfe 1997).  Observations of redshifted 21-cm absorption and
emission from DLAs indicate disk--like structures of galactic dimensions
(Briggs et al. 1989; Wolfe et al.  1992, Briggs et al. 1997), 
and Faraday-rotation observations
are consistent with the existence of micro-Gauss magnetic fields in these
systems (Wolfe, Lanzetta,
\& Oren 1992; Welter, Perry, \& Kronberg 1984; Perry, Watson, \& Kronberg
1993).

Since the total comoving density of HI in DLAs is comparable to that of stars
in the local universe, it is only natural to postulate that the cold gas out of
which most of the present population of stars had formed, was already assembled
in galaxies at $z\approx 3$ (Lanzetta et al. 1995; Wolfe et al.  1995;
Storrie-Lombardi et al. 1996a, 1996b).  However, the typical metallicity of
$Z\approx 10\% Z_{\odot}$ (Smith et al. 1996; Lu, Sargent, \& Barlow 1996, and
references therein) and dust-to-gas ratio (Fall \& Pei 1993; Pei, Fall \&
Bechtold 1991) in these systems imply that star formation was only at its
infancy at these early times. This assertion is consistent with recent
determinations of the star formation rate at high redshifts (Madau 1996;
Lowenthal et al. 1996, and references therein).  Since star formation requires
cold gas and most of the HI detected through \lya absorption lines is in damped
systems, DLAs are the natural sites for star formation at high redshifts (Fall
\& Pei 1996).  In order to unravel the star formation history of the universe,
it is of fundamental importance to probe the stars and not only the gas in
DLAs.

Direct images of distant DLAs typically reveal a luminous galactic core
which is separated by $10$--$20~{\rm kpc}$ from the line-of-sight to the
quasar (Steidel et al. 1994, 1995, 1996).  
However, any inference about the fraction of gas converted into stars in
these systems requires a prior knowledge of the Initial Mass Function (IMF)
of these stars and their formation history.  In addition, a major fraction
of the baryonic mass might reside in the outer low surface--brightness
halos of these systems.  This possibility is raised by recent microlensing
searches (Alcock et al. 1996), which indicate that a non--negligible
fraction of the massive halo of the Milky--Way galaxy might exist in the
form of Massive Compact Halo Objects (MACHOs). As a supplement to local
microlensing searches, it would be particularly interesting to examine
whether MACHOs populate the halos of galaxies at high redshifts.  While the
microlensing probability is only $\sim 10^{-6}$ in the Milky--Way halo
($\sim 10~{\rm kpc}$), its value increases up to unity in the cores of
halos at cosmological distances ($\sim 5~{\rm Gpc}$). This follows from the
linear dependence of the lensing cross--section on the observer--lens
distance for a source at infinity.  Near the center of a DLA, {\it
macrolensing} by the entire galactic potential might take place and yield
widely separated quasar images; the likelihood of macrolensing and its
generic signatures were discussed recently by Bartelmann \& Loeb (1996) and
by Perna, Loeb, \& Bartelmann (1997).

In this paper we quantify the expected microlensing probability in distant
spiral galaxies which show up as DLAs in quasar spectra. The obvious
advantage of these galactic systems is that they are selected based on
their proximity on the sky to a compact source of light in the form of a
quasar.  Detection of microlensing of the quasar can be used to probe the
stellar {\it mass} fraction of DLAs and to test whether the halos of
galaxies at high redshifts are made of MACHOs. The characteristic Einstein
radius of a solar mass lens at a cosmological distance is $\sim 5\times
10^{16}~{\rm cm}$, comfortably in between the scales of the
continuum--emitting accretion disk ($\la 10^{15}~{\rm cm}$) and the broad
line region ($\sim 3\times 10^{17}~{\rm cm}$) of a bright quasar.  This
implies that a single microlensing event would magnify the continuum but
not the broad lines emitted by the quasar. The lines would only be affected
by the macrolensing effect of the galaxy as a whole.  As a result of this
differential amplification, the equivalent width distribution of the broad
emission lines (Francis 1992) will be significantly distorted in a sample
of microlensed quasars.  Microlensing would also result in excess
variability of such quasars. In the following, we will quantify the above
signatures for a model spiral galaxy which acts as a high redshift DLA.

The existence of a population of normal disk galaxies at high redshifts
is supported by recent Keck HIRES spectra of the damped \lya absorber
towards the quasar Q 2233+1310 (Lu, Sargent \& Barlow 1997). The metal
absorption lines of the absorbing galaxy are shifted relative to its
\lya emission redshift (Djorgovski et al. 1996), indicating a rotation
velocity $\geq 200~{\rm km~s^{-1}}$ at a separation $\sim 20~{\rm kpc}$.
Here we limit our attention to galaxies of this population.
For concretness, we focus our discussion on a single spiral galaxy and
extrapolate its present--day properties back in time using the simplest
closed--box model for its baryonic content.  Based on the observed
metallicity of DLAs, we assume that the stars constitute only 10\% of the
total disk mass at $z\approx 3$, 
and that HI is gradually converted into stars until the present epoch.  The
galaxy is modeled as having three components: an old stellar bulge, a disk
made of gas and stars, and a halo.  We compute the significance of
microlensing for each of these components.  Our simplified model ignores
the diversity among different disk galaxies or DLAs and focuses on a
typical massive spiral galaxy.  Its only purpose is to examine whether
microlensing could be detected in this idealized case.  Recent imaging of
DLA galaxies (Le Brun et al. 1996) indicates that while some systems are
clearly spirals, others have irregular morphologies. A more diverse
treatment of this problem would involve many more free parameters, and is
beyond the scope of this paper.  Our primary objective here is to motivate
an observational search for these microlensing signatures, rather than to
explore the entire possible range of parameter space.

The outline of the paper is as follows.  In \S 2 we describe the
distributions of HI and stars in our model for a galactic disk.  These
distributions are then used in \S 3 to compute the microlensing probability
as a function of the intercepted HI column density of the disk. In \S 4 we
repeat the calculation for a halo made of MACHOs.  Sections 5 and 6
relate these results to observable signatures.  Finally,
\S 7 summarizes our main conclusions.

\section{Distributions of HI and Stars In the Model Disk Galaxy}

The cross--section for gravitational lensing of a high--redshift
source ($z_s\geq 2$) peaks at intermediate redshifts,
$z\sim$0.2--$0.8$ (see, e.g.  Fig. 1 in Refregier \& Loeb 1996).  In
this redshift range, there is compelling observational evidence
indicating that disk galaxies have not accreted large quantities of
matter. The evidence includes the low level of lopsidedness in spiral
galaxies (Zaritzky \& Rix 1996), the thinness and coldness of the
stellar distribution in galactic disks (T\'oth \& Ostriker 1992), the
small scatter in the Tully-Fisher relation (Eisenstein \& Loeb 1996),
the similarity between the kinematics at a given luminosity of
galaxies out to $z\sim 1$ and local galaxies (Vogt et al. 1996), and
the statistics of multiply imaged quasars (Mao \& Kochanek 1994; Rix
et al.  1994).  We therefore anchor our modeling to the current
properties of a typical galactic disk, and evolve the gaseous mass
fraction of its disk based on a simple closed box model.

Observations of local spiral galaxies show that their HI gas is generically
distributed in an exponential disk with a radial face--on column--density
profile (Broeils \& Von Waerden 1994),
\begin{equation}
N^{\rm face-on}(R,{\rm now}) = N_0 \exp(-R/R_{\rm g}),
\label{eq:5}
\end{equation}
where $N_0$ is the central column density and $R_{\rm g}$ is the scale
radius of the HI disk. For our model galaxy we adopt $N_0\approx
10^{21}{\rm cm}^{-2}$, which is the typical value observed in nearby
galaxies.

The surface mass density of stars in a nearby spiral galaxy can be modeled as 
the sum of an exponential disk and a bulge--described by a de
Vaucouleur law,
\begin{equation}
\Sigma_\star^{\rm face-on}(R,{\rm now}) = \Sigma_0 \exp(-R/R_{\rm s}) +
\Sigma_{\rm e}\exp\{-7.67[(R/R_e)^
    {\frac{1}{4}}-1]\},
\label{eq:7}
\end{equation}
where $R$ is the radial coordinate on the face of the disk.
The central mass density of stars in the disk, $\Sigma_0$, can be
calibrated from the characteristic surface brightness observed in
galactic disks $I_0\approx 21.65\mu_{B}$, where $\mu_B$ is
$B\;{\rm mag}\;{\rm arcsec}^{-2}$ (Freeman 1970). 
We convert the surface brightness to mass density assuming a mass to light
ratio of $5$ in solar units (Broeils \& Courteau 1996). 
The bulge surface density at $R=R_{\rm e}$,
$\Sigma_ {\rm e}$, can be related to $\Sigma_0$ and the total disk and
bulge luminosities, $L_{\rm d}$ and $L_{\rm b}$, through the relation
$L_{\rm d}/L_{\rm b}=0.28(R_{\rm s}/R_{\rm e})^2 (\Sigma_0/\Sigma_{\rm
e})$. This relation results from an integral of the corresponding surface
brightnesses over the area of the disk. For typical values of $L_{\rm
d}/L_{\rm b}\approx 1.3$ and $R_{\rm e}/R_{\rm s}\approx 0.4$ 
(Freeman 1970; Burstein 1979),
we get $\Sigma_{\rm e}/\Sigma_0 \approx 1.5$. The scale radius of the stars
$R_{\rm s}$ is strongly correlated with the scale radius of the gas $R_{\rm
g}$. Figure 8b in Broeils \& van Woerden (1994) illustrates that the HI
surface densities of different galaxies fall off with very similar profiles
when the radius is expressed in units of the optical radius $R_{25}$ of the
25 mag ${\rm arcsec}^{-2}$ isophote in the $B$ band.
With $I_0\approx 21.65\mu_{\rm B}$, we derive $R_{25}\approx
3.1R_{\rm s}$ and from the slopes of the profiles in Broeils
\& van Woerden (1994) we estimate $R_{\rm g}\approx 0.8R_{25}$ so that $R_{\rm
s}\approx 0.4R_{\rm g}$. For our model galaxy we adopt the typical value of
$R_{\rm g}=10$kpc.  Our results for microlensing by stars in the disk are
not very sensitive to the particular choice of $R_{\rm g}$ and mostly depend  
on the adopted ratio between $R_{\rm g}$ and $R_{\rm s}$, for which the 
observed scatter is small.

We evolve the model galaxy back in time as a closed system.  We assume that
the model disk was already assembled at $z\approx 3$ and that its HI
content evolved only as a result of star formation; this assumption is
consistent with the fact that the comoving HI density of DLAs at $z\approx
3$ is comparable to the mass density of stars in the local universe.  As
nearby galaxies exhibit a much older stellar population in their bulges
than in their disks, we keep the bulge properties fixed in time and evolve
only the stellar mass fraction of the disk with redshift for $z\la 3$.
Based on the observed metallicity of DLAs (Pettini et al. 1994, Pettini et
al. 1996) we assume that at a redshift $z\approx3$ the mass fraction of
stars in the disk was about a tenth of its present value.  Under these
assumptions, equation~(\ref{eq:5}) is generalized to
\begin{equation}
N^{\rm face-on}(R,z) = N_0 e^{-R/R_{\rm g}} + \Delta N_0(z) e^{-R/R_{\rm
s}},
\label{eq:9}
\end{equation}
where in the closed box model the HI and stellar surface densities obey
\begin{equation}
 \Sigma_{\rm HI}^{\rm face-on}(R,z) + \Sigma_{\star}^{\rm face-on}(R,z) = {\rm
const} 
\label{eq:10}
\end{equation}
with $\Sigma_{\rm HI}=m_{\rm H} N$, and therefore
\begin{equation}
\Delta \Sigma_{\rm HI}(R,z)= \Sigma_{\star,\rm disk}(R,0)-
\Sigma_{\star,\rm disk}(R,z).
\label{eq:11}
\end{equation}
Here $m_{\rm H}$ is the mass of a hydrogen atom, and the subscript
$\{\star,{\rm disk}\}$ refers to the disk stars only.  In the context of
the closed box model, the gas out of which the disk stars formed must have
had the same radial distribution as those stars have today, and so
equation~(\ref{eq:11}) can be regarded as a condition on the corresponding
central densities. We constrain the redshift evolution of $\Sigma_{\rm HI}$
by the condition $\Sigma_{\star,\rm disk}(z=3) \approx 10\%
\times\Sigma_{\star,\rm disk}(z=0)$, and by the observed redshift evolution
of the HI content of DLAs (Wolfe et al.  1995; Storrie--Lombardi et al.
1996a).  Both of these constraints are reasonably satisfied in the redshift
interval $0\la z\la 3$, if one uses a power--law redshift evolution of
the form, $\Delta\Sigma_{\rm HI}(R,z)=m_{\rm
H}{\tilde N}z^{\delta}\exp(-R/R_{\rm s})$, with $\delta\approx 6$ and
${\tilde N}=10^{20}{\rm cm}^{-2}$.  Based on equation~(\ref{eq:11}), we then
derive
\begin{equation}
\Sigma_{\star,\rm
disk}(R,z)=[\Sigma_{0} - m_{\rm H}\Delta N_0(z)]
\exp(-R/R_{\rm s})\equiv \Sigma^\prime_\star(z)\exp(-R/R_{\rm s}) ,
\label{eq:11new}
\end{equation}
and thus the stellar surface density profile is given by
\begin{equation}
\Sigma_\star^{\rm face-on}(R,z) 
= \Sigma'_\star(z) \exp(-R/R_{\rm s}) + \Sigma_{\rm e}
\exp(-7.67[(R/R_e)^{\frac{1}{4}}-1]).
\label{eq:14}
\end{equation}

Equations~(\ref{eq:9}) and ~(\ref{eq:14}) describe the radial profiles for
a face--on disk. If the normal to the 
disk is inclined by an angle $\theta$ with respect
to the line of sight, then the observed distributions are changed to
\begin{equation}
N(R,\theta,z) ={\left[ N_0 \exp({-R/R_{\rm g}}) 
       + \Delta N_0(z) \exp(-R/R_{\rm s})\right]
\over {\cos\theta}}.
\label{eq:15}
\end{equation}
and      
\begin{equation}
\Sigma_\star(R,\theta,\phi, z) = \Sigma_{\star,\rm disk} +
\Sigma_{\star,\rm bulge}=
\frac{\Sigma'_\star(z)e^{-R/R_{\rm s}}}{\cos\theta}
 + \Sigma_{\rm e} e^{-7.67[(b/R_e)^{\frac{1}{4}}-1]},
\label{eq:16}
\end{equation}
where $b=R\sqrt{1-\sin^2\theta\cos^2\phi}$ is the impact parameter and
$\phi$ is the azimuthal coordinate on the sky of the line-of-sight relative
to the center of the galaxy. In this derivation we have assumed, for
simplicity, a
razor--thin disk and a spherical bulge.

\section{Microlensing by Stars in the Disk and the Bulge}

For a single galaxy, the optical depth to microlensing is given by
the dimensionless surface mass density of its stars 
\begin{equation}
\kappa_\star\equiv\frac{\Sigma_\star}{\Sigma_{\rm crit}},
\label{eq:18}
\end{equation}
as normalized by the critical surface density, 
\begin{equation}
\Sigma_{\rm crit}=\frac{c^2 D_{\rm s}}{4\pi G D_{\rm d}D_{\rm ds}}
\label{eq:sigcrit}.
\end{equation}
Here $G$ is the gravitational constant, $c$ is the speed of light, and
$D_{\rm d}$, $D_{\rm s}$ and $D_{\rm ds}$ are the angular diameter
distances between the observer and the lens, the observer and the source
and lens and the source, respectively (Schneider, Ehlers, \& Falco 1992).
Throughout the paper we assume $H_0=50~{\rm km~s^{-1}~Mpc^{-1}}$,
$\Omega_0=1$, and a quasar redshift of $z_{\rm s}=4$.

We would like to find the probability distribution for the microlensing
optical depth, $P(\kappa_{\star})$, 
given a particular observed value for the HI
column density in our model galaxy.  We assume random inclinations of the 
galactic disk relative to the line-of-sight.
Given the HI column density in equation~(\ref{eq:15}),
the inclination angle is restricted by the condition $\theta\geq\theta_0$,
where
\begin{equation}
\cos\theta_0\equiv\gamma_0={\rm min}\left(1, \frac{N_0 + 
\Delta N_0(z)}{N}\right).
\label{eq:20}
\end{equation}
The uniform probability distribution for the cosine of the inclination
angle is then
\begin{equation}
P(\cos\theta) d \cos\theta=\frac{1}{\gamma_0} d\cos\theta~~~{\rm for}~~
0\leq\cos\theta\leq\gamma_0.
\label{eq:21}
\end{equation}

Given an inclination angle, $\theta$, the microlensing optical depth
$\kappa_\star$ can be obtained from equations~(\ref{eq:16}) and
(\ref{eq:18}). The value of $\kappa_\star$ depends on the observed column
density $N$ through the radius $R$ intercepted by the line-of-sight in the
plane of the disk; this radius is in turn obtained by inverting equation~
(\ref{eq:15}).  Given an observed HI column density $N$, the probability
distribution for the optical depth $\kappa_\star$ is
\begin{eqnarray}
P(\kappa_\star|N)&=&\frac{1}{2\pi\gamma_0}
\int_0^{2\pi}d\phi\left|\frac{d\cos\theta}{d\kappa_\star}\right|\nonumber\\
&=&\frac{1}{2\pi\gamma_0}\int_0^{2\pi}d\phi\left|-1.975\kappa_{\rm b}
b^{-\frac{3}{4}}R_e^{-\frac{1}{4}}
\frac{db}{d\cos\theta}-\frac{\kappa_{\rm d}}{{\cos}^2\theta}\left
(1 + \frac{\cos\theta}{R_{\rm s}}\frac{dR}{d\cos\theta}\right)\right|^{-1}
\label{eq:prob}
\end{eqnarray}
with
\begin{equation}
\frac{dR}{d\cos\theta}=-\frac{N}{R_g^{-1}N_0 \exp(-R/R_g)+
R_s^{-1} \Delta N_0(z) \exp(-R/R_s)},
\label{eq:dr}
\end{equation}
where $\kappa_{\rm b}=\Sigma_{\rm bulge}/\Sigma_{\rm crit}$ and
$\kappa_{\rm d}=\Sigma_{\rm disk}/\Sigma_{\rm crit}$.
Note that $P(\kappa_{\star}|N)$  depends on the absorber redshift
 through equations~(\ref{eq:18})-(\ref{eq:20}) and
(\ref{eq:dr}). 

\begin{figure}[t]
\centerline{\epsfysize=5in\epsffile{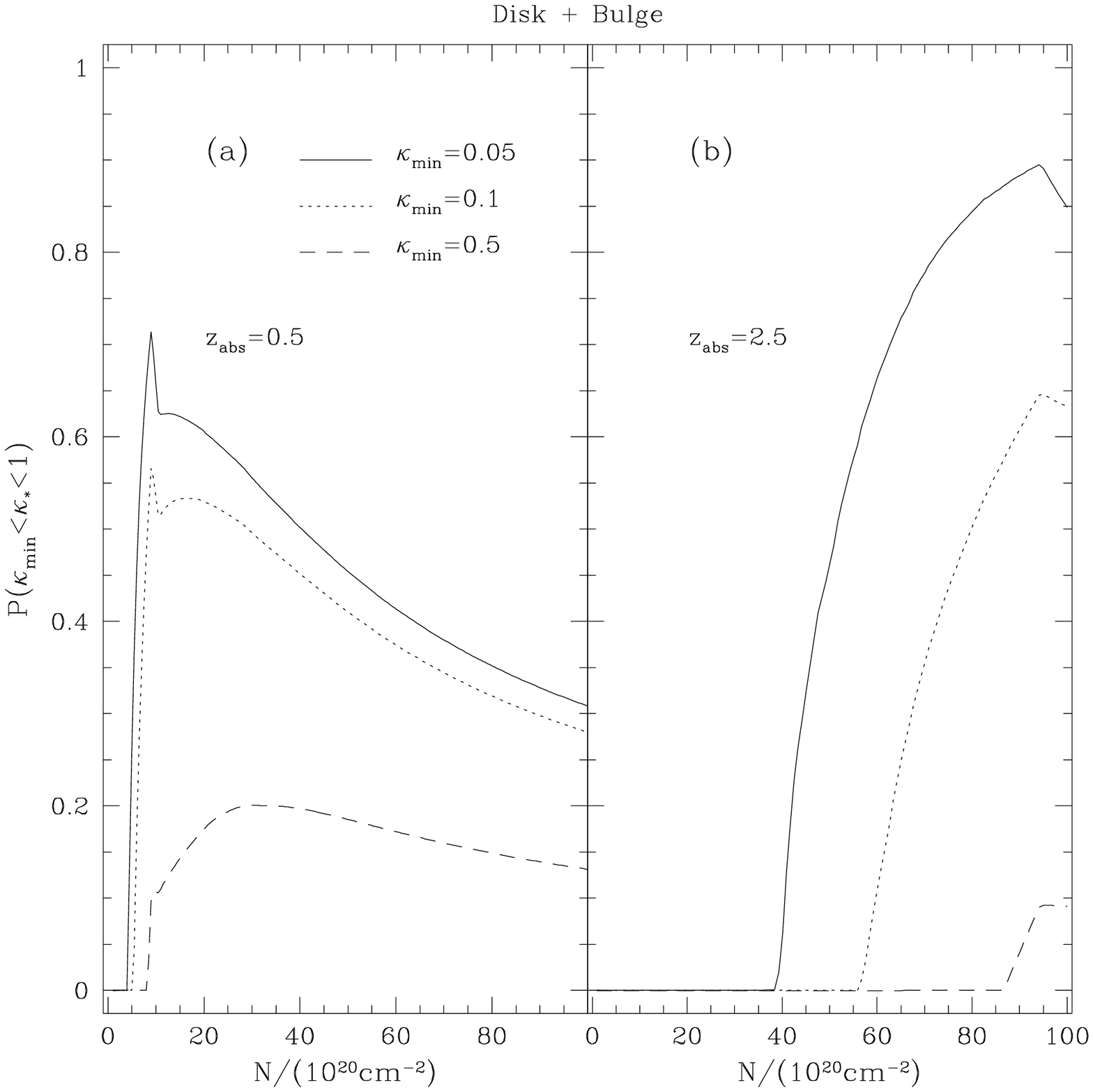}}
\caption{Probability of finding the microlensing optical depth 
$\kappa_{\star}$ in the range between $\kappa_{\rm min}$ and 1 as a
function of the observed HI column density in our model galaxy.  Results
are shown for an absorber (lens) redshift $z_{\rm abs}=0.5$ in panel (a)
and $z_{\rm abs}=2.5$ in panel (b). In both panels we show three curves for
$\kappa_{\rm min}=0.05$ (solid line), $\kappa_{\rm min}=0.1$ (dotted line),
and $\kappa_{\rm min}=0.5$ (dashed line).
The source redshift is fixed at $z_{\rm s}=4$.}
\label{fig:1}

\end{figure}

If in total $\kappa\ga 1$, then the galaxy as a whole produces multiple
images.  Under these circumstances there could still be microlensing, as
observed in the lens Q2237+0305 (Irwin et al. 1989; Wambsganss,
Paczy\'nski, \& Schneider 1990; Rauch \& Blandford 1991; Racine 1991).
However, in this case the observed spectrum will not show a single column
density since different quasar images intercept the disk at different
impact parameters and therefore probe different HI column densities. The
\lya absorption feature will then be composed of multiple troughs, 
each having a depth proportional to the magnification of its corresponding
image and a width scaling as the square root of the HI column density
probed by this image (Bartelmann \& Loeb 1996; Loeb 1997). Since the
interpretation of $N$ is not unique in this case and since most DLAs with
$N\ga 10^{20}~{\rm cm^{-2}}$ are not macrolensed (Bartelmann \& Loeb 1996),
we restrict our attention to values of $\kappa_\star\la 1$.
Figure~\ref{fig:1} shows the probability at a given value of $N$ for
obtaining a value of $\kappa_\star$ smaller than unity but higher than some
threshold $\kappa_{\rm min}$, $P(\kappa_{\rm min}<\kappa_\star< 1
|N)=\int^1_{\kappa_{\rm min}}d\kappa_\star P(\kappa_\star|N)$.  The
different panels correspond to different redshifts of our model galaxy;
panel (a) refers to $z_{\rm abs}=0.5$ and panel (b) corresponds to $z_{\rm
abs}=2.5$.  For a low absorber redshift, the probability of having a
microlensing optical depth $\kappa_\star>\kappa_{\rm min}=0.05$ is
negligible at low column densities ($N\la 3\times 10^{20}{\rm cm}^{-2}$),
but becomes significant above $10^{21}{\rm cm}^{-2}$. For a high absorber
redshift, the probability of having $\kappa_\star>0.05$ becomes significant
at yet higher column densities, $N\ga$4--9$\times 10^{21}{\rm cm}^{-2}$.
This is due to the increase in $\Sigma_{\rm crit}$ as the lens gets closer
to the source.  In order to reach the same values of $\kappa_\star$ at high
redshifts, it is necessary that the line-of-sight to the quasar will pass
closer to the center of the absorber, where the HI density is higher.  The
drop in the probabilities at yet higher HI densities is due to the 
increase in the probability of finding systems with $\kappa > 1$ (which we
exclude from our analysis).

\section{Microlensing by MACHOs}

We now consider the possibility that in addition to the stars in the disk
and the bulge, there is a population of compact objects which dominate the
mass of the halo in our model galaxy.

We model the spherically--averaged mass distribution of the entire galaxy
as a singular isothermal sphere (SIS) with a one--dimensional velocity
dispersion $\sigma$,
\begin{equation}
M_{\rm SIS}(r)=\frac{2{\sigma}^2}{G}r ,
\label{eq:22}
\end{equation} 
where $r$ is the radial spherical coordinate and $\sigma$ the velocity
dispersion for which we adopt a typical value of 170 km/sec. The SIS model
yields a flat rotation curve, as observed in nearby spirals.  We
assume that the total mass of the galaxy is composed of the disk of gas and
stars [cf.  Eqs.~ (\ref{eq:9}) and (\ref{eq:14})] and a spherical
component (denoted by SP) which includes both the bulge and the halo,
\begin{equation}
M_{\rm SIS}(r)=M_{\rm SP}(r)+M_{\rm HI,disk}(r)+M_{\star,\rm disk}(r).
\label{eq:24}
\end{equation}
The HI and stellar masses out to a radius $R=r$ on the face of the disk
are found by integrating their respective surface densities,
\begin{eqnarray}
M_{\rm HI,disk}(r)
&=&2\pi\int^r_0\Sigma_{\rm HI}^0\exp(-R/R_{\rm g})RdR\nonumber\\
&=&2\pi\Sigma_{\rm HI}^0{R_{\rm g}}^2\left[1-\exp(-r/R_{\rm g})\left(1+
    \frac{r}{R_{\rm g}}\right)\right]
\label{eq:mHI}
\end{eqnarray}
with $\Sigma_{\rm HI}^0=m_{\rm H}N_0$ and 
\begin{equation}
M_{\star,\rm disk}(r)=2\pi\Sigma_{0}{R_{\rm s}}^2\left[1-\exp(-r/R_{\rm
s})\left(1+ \frac{r}{R_{\rm s}}\right)\right].
\label{eq:m*}
\end{equation}
The mass density of the spherical component (bulge plus halo)
can be derived from equation~(\ref{eq:24}) through the relation,
\begin{equation}
\rho_{\rm SP}(r)=\frac{1}{4\pi r^2}\frac{dM_{\rm SP}}{dr}
=\frac{\sigma^2}{2\pi{G}r^2}-\frac{1}{2r}\left[\Sigma_{\rm HI}^0 \exp
(-r/R_{\rm g})+\Sigma_{0} \exp(-r/R_{\rm s})\right].
\label{eq:ro}
\end{equation}
The surface density as a function of the impact parameter on the sky $b$
is then found by integrating this mass density
along the line of sight $\Sigma(b)=\int_{-\infty}^{\infty}d\zeta\rho(b,\zeta)$
with $r=(b^2+\zeta^2)^{\frac{1}{2}}$. This yields
\begin{equation}
\Sigma_{\rm SP}(b)=\frac{\sigma^2}{2{G}b}
-\frac{\Gamma(1/2)}{\sqrt{\pi}}\left[\Sigma_{\rm HI}^0\;{\rm K_0}
\left(\frac{b}{R_{\rm g}}\right)+\Sigma_{0}\;{\rm K_0}\left(
\frac{b}{R_{\rm s}}\right)\right],
\label{eq:25}
\end{equation}
where ${\rm K_0}$ is the modified Bessel function of the zeroth order.
The optical depth to microlensing is then given by 
\begin{eqnarray}
\kappa (R,\theta,\phi)&=& \kappa_{\rm SP}+\kappa_{\star,\rm disk}\nonumber\\
       &\equiv& \frac{1}{\Sigma_{\rm crit}}\left(
\Sigma_{\rm SP}(b)+\frac{\Sigma_{0}\exp(-R/R_{\rm s})}
    {\cos\theta}\right),
\label{eq:28}
\end{eqnarray}
where, as before, $\theta$ is the inclination of the normal to 
the disk relative to the
line-of-sight, $\phi$ is the azimuthal coordinate on the sky,
$R$ is the radial coordinate in the plane of the disk, and
$b=R\sqrt{1-\sin^2\theta\cos^2\phi}$.
The probability distribution of $\kappa_\star$ for an
observed HI column density $N$, can be calculated again as in
equation~(\ref{eq:prob}),
\begin{eqnarray}
&&P(\kappa_\star|N)=\frac{1}{2\pi\gamma_0}\int_0^{2\pi}d\phi\nonumber\\
&\times&
\left|\frac{db}{d\cos\theta}\left[-\frac{\kappa^0_{\rm SIS}}{b^2}+
\frac{\kappa_{\star,\rm disk}^0}{R_{\rm s}}\;{\rm K_1}\left(\frac{b}{R_{\rm s}}\right)+
\frac{\kappa_{\rm HI}^0}{R_{\rm g}}\;{\rm K_1}\left(\frac{b}{R_{\rm g}}
\right)\right]-\frac{\kappa_{\star,\rm disk}}{{\cos}^2\theta}\left(1+
\frac{\cos\theta}{R_{\rm s}}\frac{dR}{d\cos\theta}\right)\right|^{-1}
\label{eq:probh}
\end{eqnarray}
where ${\rm K_1}$ is the modified Bessel function of the first order, and
$dR/d\cos\theta$ is given in equation~(\ref{eq:dr}).  

\begin{figure}[t]
\centerline{\epsfysize=5in\epsffile{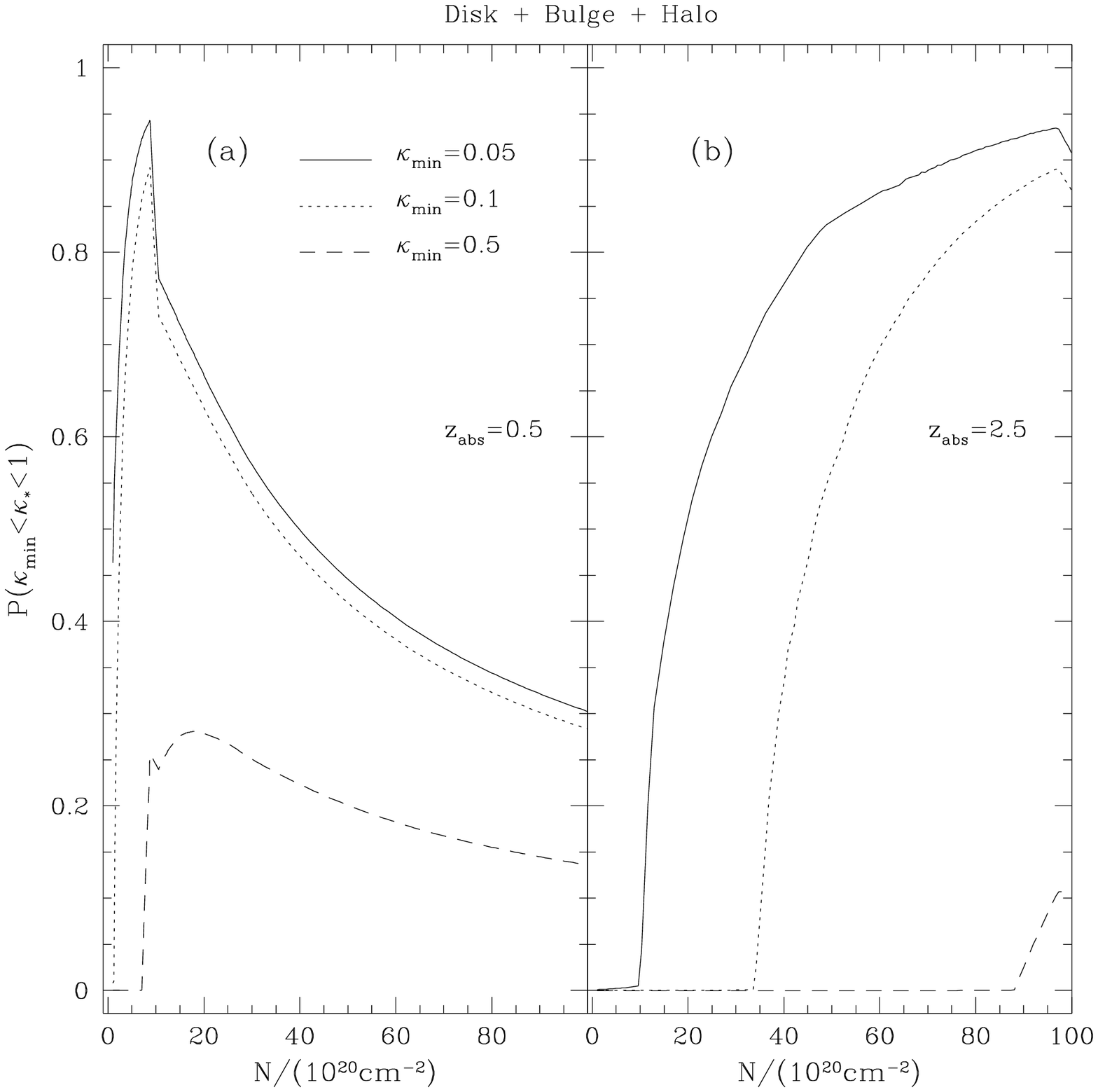}}
\caption{Same as Figure 1, but including a halo of MACHOs 
in addition to the bulge and the disk stars.  The total microlensing 
optical depth is 
$\kappa_\star=\kappa_{\star,{\rm disk}} +\kappa_{\rm SP}$
[cf. Eqs. (22) and (23)].
}
\label{fig:2}
\end{figure}

Figure~\ref{fig:2} shows the probability for obtaining a value of
$\kappa_\star$ smaller than unity but higher than some threshold
$\kappa_{\rm min}$ at either a low absorber redshift [$z_{\rm
abs}=0.5$ in panel (a)] or a high absorber redshift [$z_{\rm abs}=2.5$
in panel (b)]. As evident from comparing Figures 2 and 1, the
existence of MACHOs enhances considerably the microlensing probability
in DLAs which are not {\it macro}--lensed. This is especially true 
for absorbers with lower column densities. 

\section{Observational Signatures of Microlensing}

The microlensing probability derived in the previous sections can be
related to two observable signatures: (i) distortion in the
equivalent--width distribution of the broad emission lines of microlensed
quasars (Canizares 1982), and (ii) excess variability of microlensed
quasars (see, e.g. Wambsganss \& Kundic 1995). In the following we quantify
both of these effects as a function of the observed HI column--density for
our model galaxy.

\subsection{Equivalent--Width Distribution of Microlensed Quasars}

Quasars can be significantly affected by microlensing only if the size
of their emission region is smaller than the projected size $r_{\rm
E}$ of the lensing zone, i.e. the ``Einstein radius'' of the lensing
star.  The maximum magnification of a finite circular source of radius
$r_s$ and uniform surface brightness is given by [Schneider et al.
(1992)]
\begin{equation}
\mu_{\rm max}=\sqrt{1+4(r_{\rm E}/r_s)^2}\; .
\label{eq:mumax}
\end{equation}  
At cosmological distances, the lensing zone of a star of mass $M_{\rm
star}$, has a characteristic scale of $\sim 5\times 10^{16}~(M_{\rm
star}/M_\odot)^{1/2}~{\rm cm}$.  In comparison, the optical continuum
emission in quasars is believed to originate from a compact accretion
disk.  The UV bump observed in quasar spectra is often associated with
thermal emission from an accretion disk with a surface temperature
$T_{\rm disk}\equiv 10^5 T_5~{\rm K}$, where $T_5\sim 1$ (e.g. Laor
1990), and so the scale of the disk emission region must be $\sim
10^{15}~{\rm cm}~T_5^{-2} L_{46}^{1/2}$, where $L_{46}$ is the
corresponding luminosity of the quasar in units of $10^{46}~{\rm
erg~s^{-1}}$.  Thus, for lens masses $M_{\rm star}\gg 10^{-3}M_\odot$,
the continuum source is much smaller than the lensing zone and could
therefore be magnified considerably.  This expectation is indeed
confirmed in the nearby lens of Q2237+0305, where variability due to
microlensing has been observed (Wambsganss et al. 1990; Rauch \&
Blandford 1991; Racine 1991; see also Gould \& Miralda-Escud\'e 1996).
However, reverberation studies of the time lag between variations in
the continuum and the line emission in active galactic nuclei indicate
that the broad emission lines of quasars originate at a distance of
$\sim 3\times 10^{17}~{\rm cm}~L_{46}^{1/2}$ (Peterson 1993; Maoz
1996). For a solar mass lens with luminosity $10^{46}$ erg cm$^{-2}$,
use of Equation~(\ref{eq:mumax}) shows that
the maximum magnification differs from unity by only a few percent.
This implies that microlensing of the broad line region by a single
star can be safely ignored.
The broad line region would only be macrolensed by a factor
$\langle\mu\rangle$ due to the average effect of all the stars.  As a
result of the difference in amplification between the lines and the
continuum, the equivalent width of the lines will change during a
microlensing event.  For compact objects which are uniformly distributed
throughout the universe, the microlensing optical depth is a function of
redshift and should therefore result in an apparent evolution of the
equivalent--width distribution of quasars.  The lack of such evolution in
existing quasar samples was therefore used to set an upper limit on the
mean cosmological density of stars and MACHOs in the universe (Dalcanton et
al.  1994). Nemiroff (1988) examined the effect of microlensing on the
shape of the broad emission lines, and Hawkins (1996) addressed the
potential to detect cosmologically distributed microlenses based on quasar
variability data.
Here we propose to look for microlensing signatures through a comparison
between the equivalent width distribution of quasars with DLAs and those
without.  Since quasars with DLAs are selected based on their small
projected separation from foreground galaxies, they are more likely to be
lensed by stars or MACHOs than the rest of the quasar population.  The
latter approach is likely to have a larger signal-to-noise ratio if most of
the stellar objects in the universe are grouped into galaxies.  Since DLAs
appear only for a small fraction of all quasars, $\sim 10\%(N/10^{20}~{\rm
cm^{-2}})^{-1/2}$, it should be straightforward to obtain an accurate
equivalent--width (EW) distribution for a control sample of quasars which
are not microlensed.

In order to find the effect of microlensing on the EW distribution of
quasar emission lines, it is necessary to calculate the magnification
probability as a function of $\kappa_\star$.  This probability function was
derived in the literature through extensive numerical simulations (Witt
1993; Wambsganss 1990; Lewis et al. 1993; Kundic \& Wambsganss 1993;
Wambsganss \& Kundic 1995).  In particular, the simulations of Rauch et al.
(1992) demonstrated that the probability distribution of magnifications for
a point source seen through a field of stars can be reasonably well
approximated by the analytic expression proposed by Peacock (1986)
\begin{equation}
P(\mu,\chi)d\mu={\rm e}^{f(\mu,\chi)}\frac{df}{d\mu}
\label{eq:pmu}
\end{equation}
where
\[ f(\mu)=2\chi[1-\mu(\mu^2-1)^{-1/2}]\;\;\;\;\; {\rm and}\;\;\;\;\; 
df/d\mu=2\chi(\mu^2-1)^{-3/2} \]
and the parameter $\chi$ is related to the dimensionless surface mass
density $\kappa_\star$ through the equation
\begin{equation}
 \langle\mu\rangle\equiv\int_1^{\infty}\mu P(\mu,\chi)d\mu=
\frac{1}{(1-\kappa_\star)^2}. 
\label{eq:avemu}
\end{equation}
The probability distribution of magnifications for a given value of the
observed HI column density $N$ can be obtained by combining the probability
distribution for $\kappa_\star$ at a given $N$ [cf. Eqs.~(\ref{eq:prob}) or
~(\ref{eq:probh})] with the probability
distribution of magnifications at a given $\kappa_\star$
[cf. Eq.~(\ref{eq:pmu})], 
\begin{equation}
P(\mu|N)=\int_0^1 d\kappa_\star P(\mu|\kappa_\star)P(\kappa_\star|N).
\label{eq:44}
\end{equation}

We define ${I}_{\nu}(\lambda_0)$ to be the intensity of the quasar
continuum in the neighborhood of the wavelength $\lambda_0$ of a particular
emission line and $\Delta{I}_{\nu} (\lambda)$ to be the difference between
the total measured intensity and the continuum.  If the continuum is
magnified by a factor $\mu$, then its observed intensity changes to
$\mu{I}_{\nu}(\lambda_0)$. The extended broad line region is affected by
the combined effect of many stars, and so the intensity of the lines is
enhanced by the average magnification factor $\langle\mu\rangle$ in
equation (\ref{eq:avemu}).  Consequently, the equivalent width (EW) of the
emission line, defined as
\begin{equation}
W_{\lambda}\equiv\int\frac{\Delta{I}_{\nu}}{{I}_{\nu}(\lambda_0)}
d\lambda
\label{eq:wl}
\end{equation}
is changed by a factor $\langle\mu\rangle/\mu$, namely
\begin{equation}
W_{\lambda}=W_0\frac{\langle\mu\rangle}{\mu}
\label{eq:wnew}
\end{equation}
where $W_0$ is the intrinsic EW of the unlensed
quasar.

Even in the absence of lensing, quasars do not possess a unique EW value in
their emission lines but rather show a wide probability distribution of EW
values (Francis 1992), which we define as $P(W_0)$.  The magnification due to
lensing distorts this probability distribution.  The modified
distribution of EW values for quasars which show DLA absorption with a
column density $N$ would be,
\begin{equation}
P(W_{\lambda}|N)=\int_1^{\infty}d\mu P(\mu|N)
P\left(\frac{\mu}{\langle\mu\rangle}
W_{\lambda}\right)\frac{\mu}{\langle\mu\rangle}.
\label{eq:pwl}
\end{equation}
The observed probability distribution of EW can be well approximated as a
log--Gaussian function (see also Dalcanton et al. 1994), 
\begin{equation}
P(W_0)=\frac{1}{\sqrt{2\pi\sigma_{W}^2}W_0} \exp\{{-[\ln(W_0)-\omega]^2/
{2\sigma_{W}^2}}\},
\label{eq:pw0}
\end{equation}
where the parameters $\omega$ and $\sigma_{W}$ obtain different values for
different emission lines.  We optimize these values so as to fit the data
from the Large Bright Quasar Survey (LBQS) reported by Francis (1992).

Figures 3 and 4 illustrate the unlensed distributions (solid lines) chosen
to best represent the observational data from LBQS, part of which is shown
as squares.  The figures also show the distorted EW distribution for quasars
which are lensed by stars in the disk (dotted--dashed line), disk+bulge
(dashed line) and disk+bulge+halo (dotted line).  Figure 3 corresponds to
our model galaxy being at a low redshift, $z_{\rm abs}=0.5$ and with an
observed column density $N=5\times 10^{20}{\rm cm}^{-2}$ (Fig. 3a) or
$N=10^{21} {\rm cm}^{-2}$ (Fig. 3b).  Figure 4 shows the results
for $z_{\rm abs}=2.5$ with $N=5\times 10^{21}{\rm cm}^{-2}$ (Fig. 4a) or
$N=9\times 10^{21}{\rm cm}^{-2}$ (Fig.  4b).  Note that the area below
the curves is in some cases smaller than unity, because we restrict our
attention to systems with $\kappa_\star<1$.  In all cases, the distortion
of the EW distribution is more pronounced for galaxies with MACHOs.
The level of this distortion can therefore be used as a sensitive probe of
the composition of galactic halos at high redshifts, and by that complement
searches for MACHOs in the local universe (Alcock et al. 1996, and
references therein).

In order to estimate the minimum size of the quasar sample that allows
detection of the distortion to the EW distribution, we performed a numerical
simulation of random deviates drawn from the lensed (i.e.  distorted)
distribution and computed the ${\chi}^2$ statistic with respect to the observed
data for the unlensed distribution (Francis 1992).  We assumed that the number
of quasars with DLAs is much smaller than the number of quasars without DLAs,
which were used to define the unlensed distribution.  In this case,
${\chi}^2=\sum_i [{({\hat n}_i-n_i)^2}/{n_i}]$, where ${\hat n}_i$ is the
number of simulated events from the lensed distribution in the $i$--th bin and
$n_i$ is the number of expected events according to the unlensed distribution.
For low-redshift ($z_{\rm abs}\sim 0.5$) absorbers with halos made of MACHOs
and observed column densities $N\sim 10^{21} {\rm cm}^{-2}$, we find that a
relatively small sample of $\sim 10$ systems drawn from the lensed distribution
can be rejected as being drawn from the unlensed distribution with a ${\chi}^2$
probability of $\la 5\times 10^{-2}$. 
 
If MACHOs are absent, the number of systems required to get the same
confidence level is increased to $\sim 20$. For high redshift ($z_{\rm
abs}\sim 2.5$) absorbers with MACHOs and $N\sim 10^{21} {\rm cm}^{-2}$, the
required number of systems, $\sim 100$, exceeds the number of {\it all}
known DLAs. However, for high--redshift absorbers with column densities on
the order of $\sim 8-10\times 10^{21}~{\rm cm}^{-2}$, even a small sample
of $\sim 10$ systems could be sufficient to detect the signal at the same
confidence level.

We note that our calculations were formulated in terms of the stellar
surface density $\kappa_\star$, and therefore ignored contributions to the
magnification distribution from shear $\gamma$, or additional sources of
$\kappa$ (e.g. due to a smoothly distributed dark matter component).  
%
The associated increase in the average magnification
$\langle\mu\rangle$ would affect both the lines and the continuum.  In
addition, the finite source size would tend to reduce the expected
signal relative to our point source calculation.  In reality, the
magnification is limited to a value $\mu_{max}\sim r_{\rm E}/r_s$ for
$r_s\ll r_{\rm E}$ (see Equation~(\ref{eq:mumax}). 
%
For a stellar mass of $\sim 0.3 M_\odot$ and a continuum
emitting region of a quasar of size $\sim 10^{15}~{\rm cm}$, we find
$\mu_{max}\sim 10^2$. The contribution to $P(\mu)$ from higher values of
$\mu$ is negligible; $P(\mu\ga 100)\la 10^{-7}$ for $\kappa_\star=0.1$ and
$P(\mu\ga 100)\la 10^{-4}$ for $\kappa_\star=0.8$.

Our calculation of the EW distribution ignores any magnification bias due
to lensing, by which faint QSOs are raised above the detection limit. This
bias could affect our analysis only if the EW were to depend on the
luminosity of the source. Such a dependence, otherwise known as the
``Baldwin Effect'' (Baldwin 1977; Baldwin et al.  1988), is generally very
weak and was only marginally detected for the CIV line (Francis 1992).
By modeling EW$\propto L^{-0.1}$ , this weak
scaling results in a $\sim 1\%$ change of the EW values (due to
magnification bias) for $\kappa_\star=0.1$ and in a $\sim 5\%$ change for
$\kappa_\star=0.8$.  Thus, the magnification bias can be safely ignored in
our calculation.

\subsection{Time variability} 

The microlensing signature discussed in the previous subsection can be
detected at one point in time, as it only requires a statistical comparison
between the EW distribution of quasars which have DLAs in their spectra and
that of quasars which do not.  By monitoring a quasar over a sufficiently
long period of time, it is possible to search for yet another signature of
microlensing, namely an excess temporal variability of the quasar flux due
to the motion of the microlenses relative to the quasar.  Below we quantify
the probability for detecting this excess variability as a function of the
observed HI column density in our model galaxy.

The net motion of the stars relative to the quasar is the sum of their
motion within the galaxy and the bulk velocity of the galaxy relative to
the line-of-sight to the quasar.  Wambsganss \& Kundic (1995) have
calculated the probability distribution for observing an event with
duration smaller than the crossing time of the stellar Einstein--radius
$t_0$ as a function of $\kappa_\star$ for a population of single--mass
stars.  Given the source and lens redshifts, the value of $t_0$ depends on
the mass $M_{\rm star}$ of the lensing star and on its transverse velocity
$v$ relative to the line-of-sight,
\begin{equation}
 t_0= {\sqrt{(4GM_{\rm star}/c^2)(D_{\rm d} D_{\rm ds}/D_{\rm s})} \over v},
\label{eq:31}
\end{equation}
while an ``event'' was defined as an increase in the magnification by a
factor $\Delta m \ga 0.2$ mag.  Wambsganss \& Kundic (1995) presented the
probability distribution of $t_0$ for simulations with $\kappa_\star=0.2$
in their paper, and have kindly provided us with their results also for
$\kappa_\star=0.5$ and $\kappa_\star=0.8$ (Wambsganss
\& Kundic 1996, private communication).  In order to obtain a rough 
estimate for the probability distribution 
$P(t\leq t_{\rm max}|\kappa_\star)$, we
have interpolated these numerical results over the entire range of
$0\leq\kappa_\star<1$.  Then, by combining this distribution with
equation~(\ref{eq:prob}) or equation~(\ref{eq:probh}), we have obtained the
probability for observing a microlensing event with duration less than
$t_{\rm max}$ given that the observed HI column density is $N$,
\begin{equation}
P( t\leq t_{\rm max}|N)=\int_0^1 d\kappa_\star P( t\leq t_{\rm
max}|\kappa_\star) P(\kappa_\star|N),
\label{eq:30}
\end{equation}

The transverse velocities of stars in galaxies are typically of order
$150$--$200~{\rm km~s^{-1}}$. The characteristic transverse velocity of
galaxies due to large--scale structure can be as large as 500 km/sec
(Strauss \& Willick 1995, and references therein).  We therefore
parametrize, $v=300\times v_{300}~{\rm km/sec}$.  The simulation of
Wambsganss \& Kundic (1995) assumes a single mass for all stars.  We
estimate the numerical value of $t_0$ by averaging over a Scalo (1986) mass
function for the stars, with $\langle M^{1/2}\rangle\approx
0.63{M_{\odot}}^{1/2}$.  Using a source redshift $z_{\rm s}=4$, we then get
$t_0\approx 28 v_{300}^{-1}\;{\rm yr}$ for a lens at $z_{\rm abs}\approx
0.5$ and $t_0\approx 17 v_{300}^{-1}\;{\rm yr}$ at $z_{\rm abs}\approx
2.5$.  Note that the above time scales are averaged over the stellar mass
function.  In reality, the low mass tail of the stellar distribution could
give rise to events of much shorter duration.

Our results are shown in Figures~\ref{fig:5} (for disk+bulge) and
~\ref{fig:6} (disk+bulge+halo) for a 
low redshift absorber [panel (a)] or a high
redshift absorber [panel(b)].  The figures show the probability of
observing a microlensing event with duration less than $t_{\rm
max}=0.08t_0$ (solid line), $0.12t_0$ (dotted line), $0.16t_0$ (dashed
line), and $0.2t_0$ (dotted--dashed line).

The variable magnification during a microlensing event could be detected
through two effects:
\begin{itemize}
\item[1.] {\it Changes in the EW of the broad emission lines with time}.
A magnification factor of $\Delta m\ga 0.2$ mag
will reduce the EW by a factor $\ga 0.2$ 
over time. This change could be detected through a monitoring program of
quasar emission lines, of the type realized in reverberation studies of
quasar variability (Kaspi et al. 1996). It should be generally distinguishable
from changes due to intrinsic variability, where an increase in the 
continuum intensity is followed, after a certain time, by a corresponding
increase in the intensity of the emission lines (Maoz 1996). 

\item[2.] {\it Variability in the QSO light curve}. Excess
variability in the quasar luminosity due to microlensing could in principle
be distinguished from intrinsic quasar variability.  In the low optical
depth regime ($\kappa\ll1$), the microlenses are isolated and the light
curves they produce have generic shapes which are symmetric in time
(Paczy\'nski 1986). 
Even at higher optical depths, there should still be a statistical symmetry
between the rising and the falling parts of the light curves. Finally, as
long as the continuum emission region is much smaller than the size of the
stellar lensing zone, the excess temporal variation of the quasar should be
achromatic.
\end{itemize}

\section{Conclusions}

We have shown that the subset of all quasars which have damped
\lya absorption in their spectra due to an intervening
spiral galaxy should possess an enhanched tail in the equivalent width
distribution of their broad emission lines, and exhibit excess
variability, relative to the rest of the quasar population.  Both of
these microlensing signatures are more pronounced if the halos of DLAs
are composed of compact objects.

Previous attempts to find a redshift evolution in the equivalent--width
distribution of a large sample of quasars due to microlensing by a population
of intergalactic stars have yielded a negative result (Dalcanton et al. 1994).
In this paper we have shown that the significance of microlensing relative to
the statistical noise should be much more pronounced in a subset of all quasars
which are located behind galactic HI disks.  Using a simple closed--box model
for a spiral galaxy we have found that disk+bulge microlensing could be
detected through its imprint on the equivalent--width distortion with a
signal-to-noise ratio $\ga 2$ in a sample of $\approx 20$ damped \lya absorbers
(DLAs) at $z_{\rm abs}\la1$ with $N\sim 10^{21}{\rm cm}^{-2}$. The necessary
sample size is reduced to $\approx 10$ if the galaxy halo is composed of MACHOs
with a velocity dispersion of $\sigma\approx170$ km/sec.  The necessary sample
size should scale as $\propto \sigma^{-4}$ for other galactic systems and
therefore massive galaxies are likely to dominate the statistics. In addition,
we find that about a tenth of all quasars with DLAs are likely to show excess
variability on timescales shorter than five years (cf. Figures 5 and 6).

Unfortunately, the current sample of $\sim 80$ quasars with DLAs (Wolfe et
al. 1995) includes only several absorbers with $z\la1$ and might not be
sufficiently large to demonstrate the existence of microlensing. The number
of known DLAs could increase by an order of magnitude as a result of
spectroscopic follow--ups on the catalog of $\sim 10^5$ quasars which is
expected to be compiled by the forthcoming Sloan Digital Sky Survey (Gunn
\& Knapp 1993). Even before any absorption data is reduced, it would be
interesting to select background quasars which are projected close to
foreground galaxies and therefore are likely to be microlensed. Related
studies by Webster et al. (1988) showed an increased number of
quasar-galaxy pairs
than expected from random alignments. The enhancement in the quasar surface
density near galaxies was interpreted as macro-lensing and used to draw
conclusions about the distribution of matter around galaxies.  

Detection of the level of microlensing for quasars with damped \lya
absorption can be used to calibrate 
the mass fraction in the form of massive compact objects in galactic
halos at high redshifts. Figures 3-6 show that the microlensing signal
is enhanced when an isothermal halo made of MACHOs is added to the galactic
disk and bulge.
  Direct imaging of the DLAs can be used to infer
the projected separation between the luminous center of the absorber and
the quasar. In cases where the DLA redshift is known (e.g.  through its
\lya emission; see Djorgovski et al. 1996, and Lu et al. 1997), it might 
also be possible to infer spectroscopically the velocity dispersion of the
intervening galaxy.  When combined with the information gathered by
microlensing searches in the local universe, such studies could extend our
knowledge of the composition of galactic halos out to redshifts as high as
$z\sim 5$.

\acknowledgements

We thank an anonymous referee for insightful comments that improved the
presentation.  This work was supported in part by the NASA ATP grant
NAG5-3085 and the Harvard Milton fund (for AL) and by a fellowship from the
university of Salerno, Italy (for RP).


\newcounter{figmain}
\newcounter{figsub}[figmain]
\renewcommand{\thefigure}{\arabic{figmain}.\alph{figsub}}
\refstepcounter{figmain}
\refstepcounter{figmain}
\refstepcounter{figmain}
\refstepcounter{figsub}

\begin{figure}[t]
\centerline{\epsfysize=5.7in\epsffile{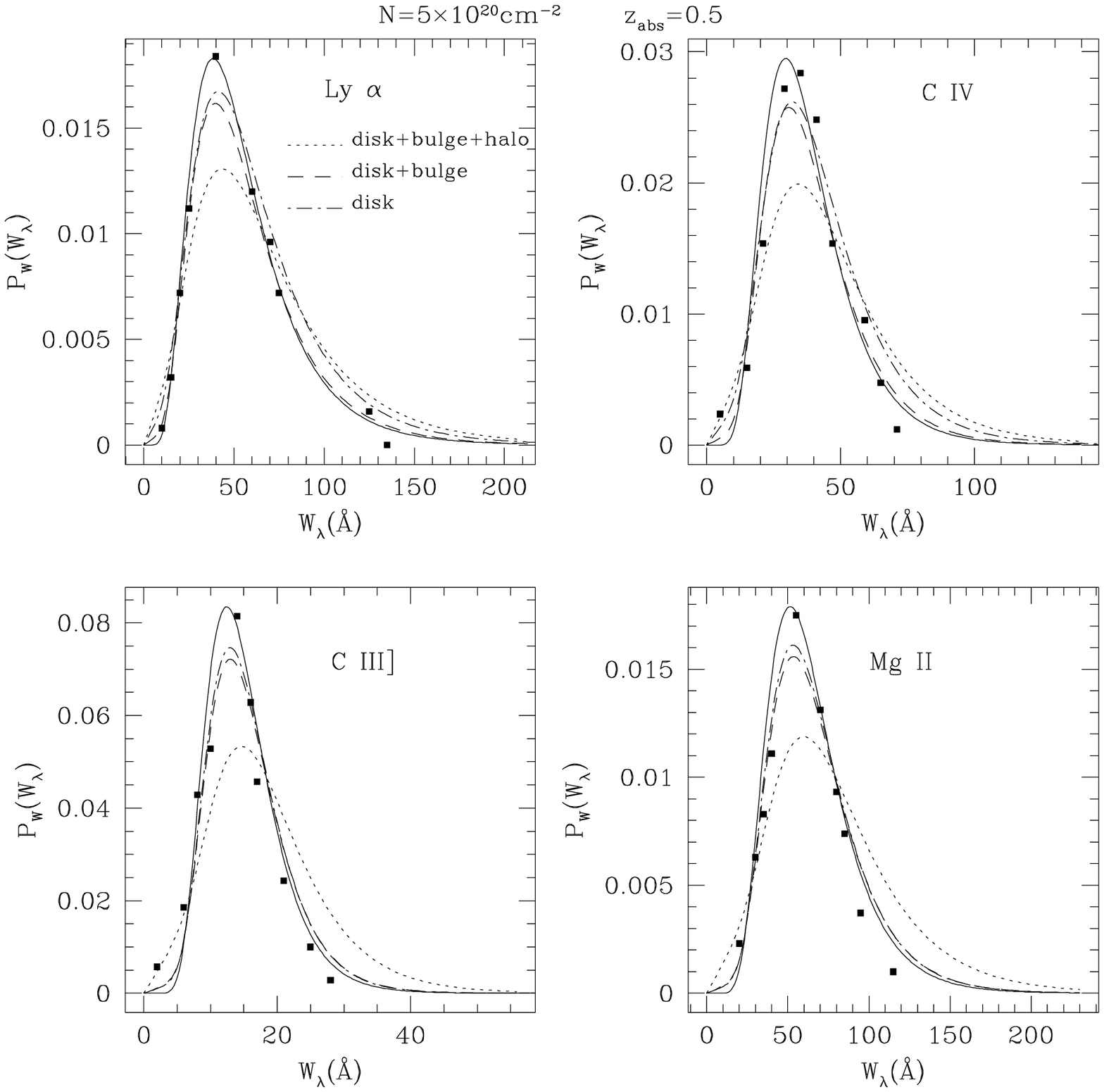}}
\caption{Effect of microlensing on the equivalent--width distribution 
of quasars with DLAs. The solid line shows the analytical log--Gaussian fit
to the observational data (squares) based on
the Large Bright QSO Survey (Francis 1992).
The other lines show the distortion of
this distribution due to microlensing
by stars in the disk (dashed-dotted line), the disk+bulge (dashed line), and
the disk+bulge+halo (dotted line) of our model galaxy.
The galaxy is assumed to be observed with an HI column density 
$N=5\times 10^{20}{\rm cm}^{-2}$ and a redshift $z_{\rm
abs}=0.5$. The source redshift is $z_{\rm s}=4$.}
\label{fig:3a}
\end{figure}

\refstepcounter{figsub}
\begin{figure}[t]
\centerline{\epsfysize=5.7in\epsffile{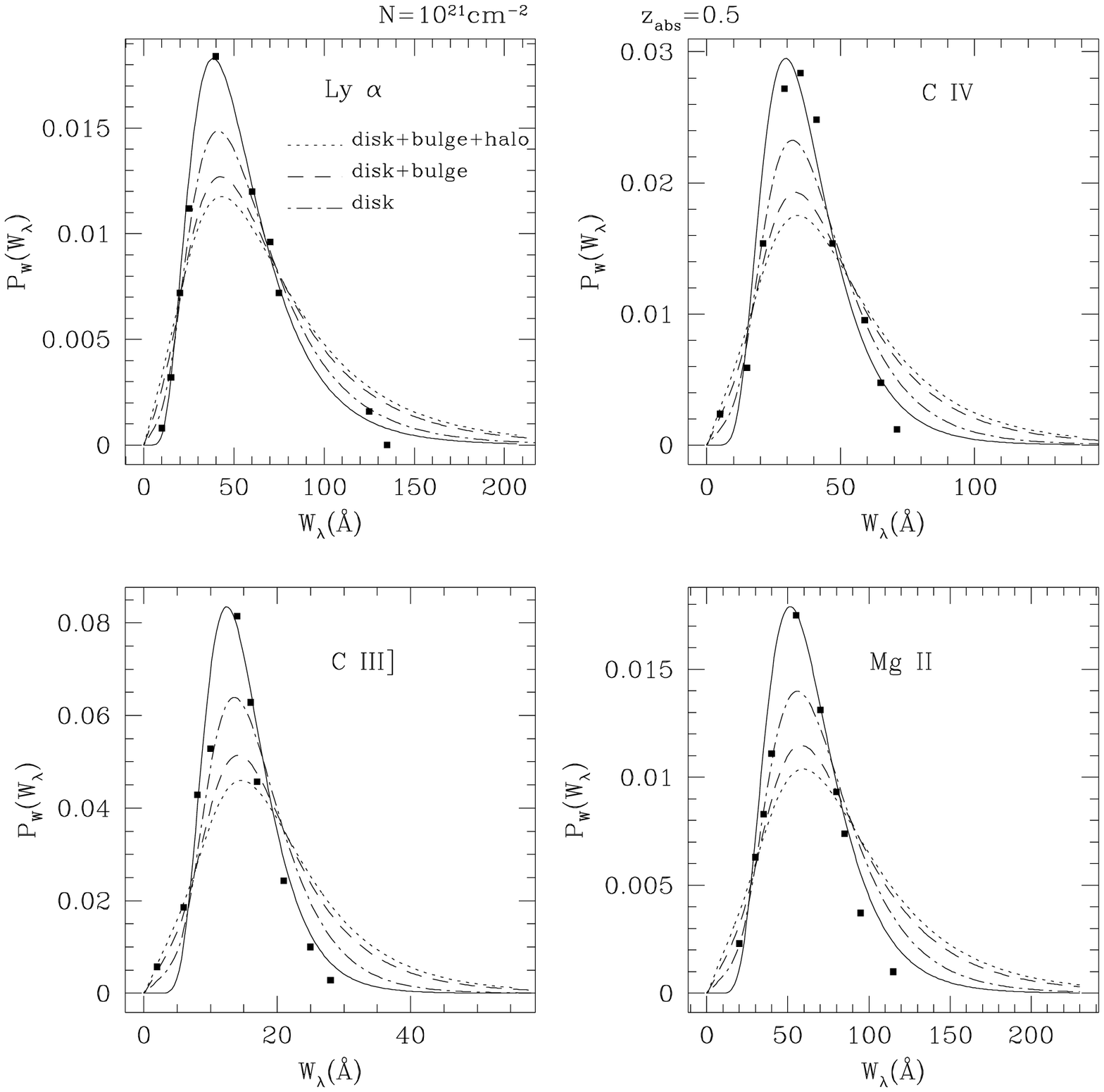}}
\caption{Same as Figure 3a, but with $N=10^{21}{\rm cm}^{-2}$.}
\label{fig:3b}
\end{figure}

\refstepcounter{figmain}
\refstepcounter{figsub}
\begin{figure}[t]
\centerline{\epsfysize=5.7in\epsffile{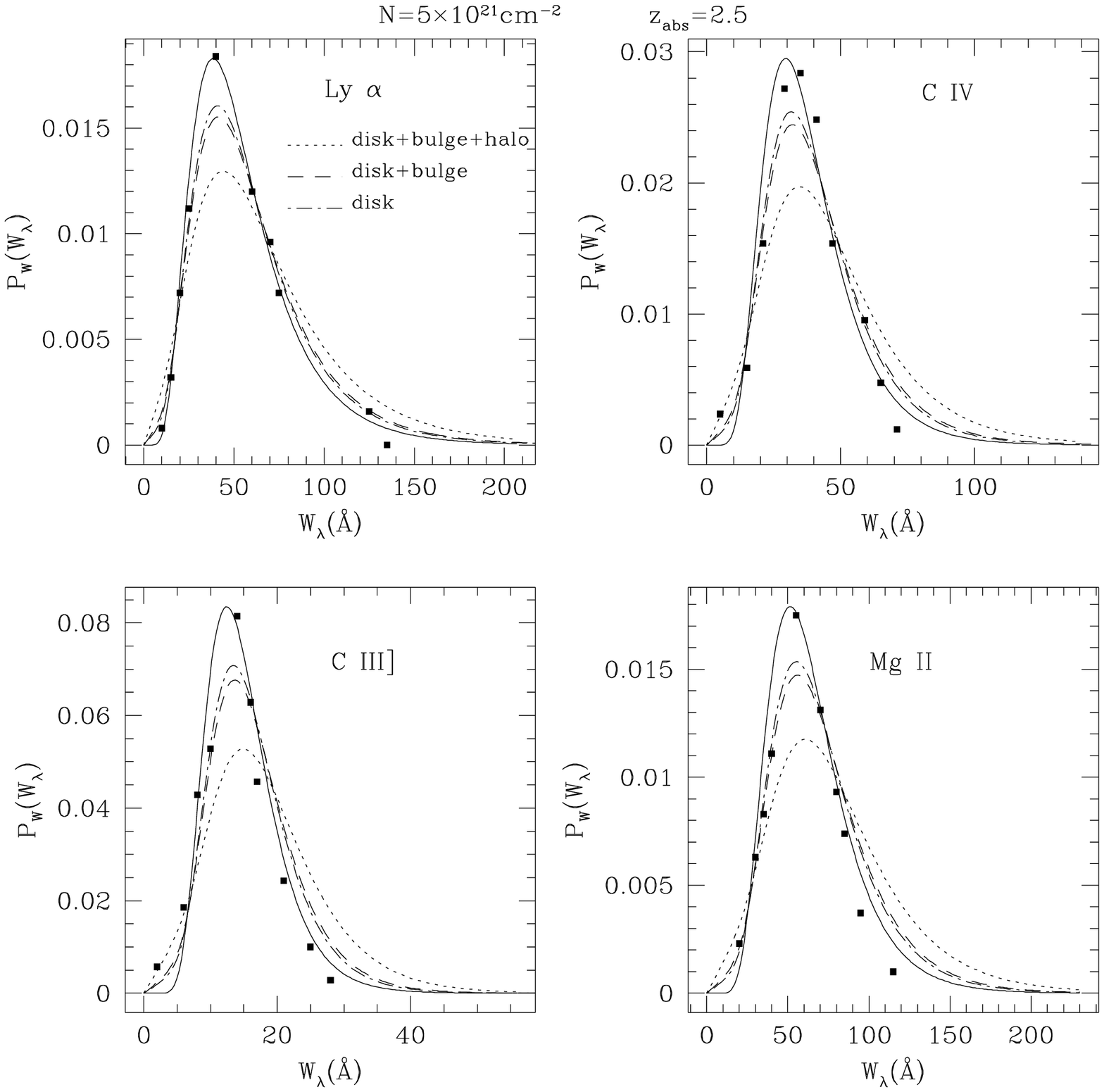}}
\caption{Same as Figure 3a, but with $z_{\rm abs}=2.5$ and 
  $N=5\times 10^{21}{\rm cm}^{-2}$.}
\label{fig:4a}
\end{figure}

\refstepcounter{figsub}
\begin{figure}[t]
\centerline{\epsfysize=5.7in\epsffile{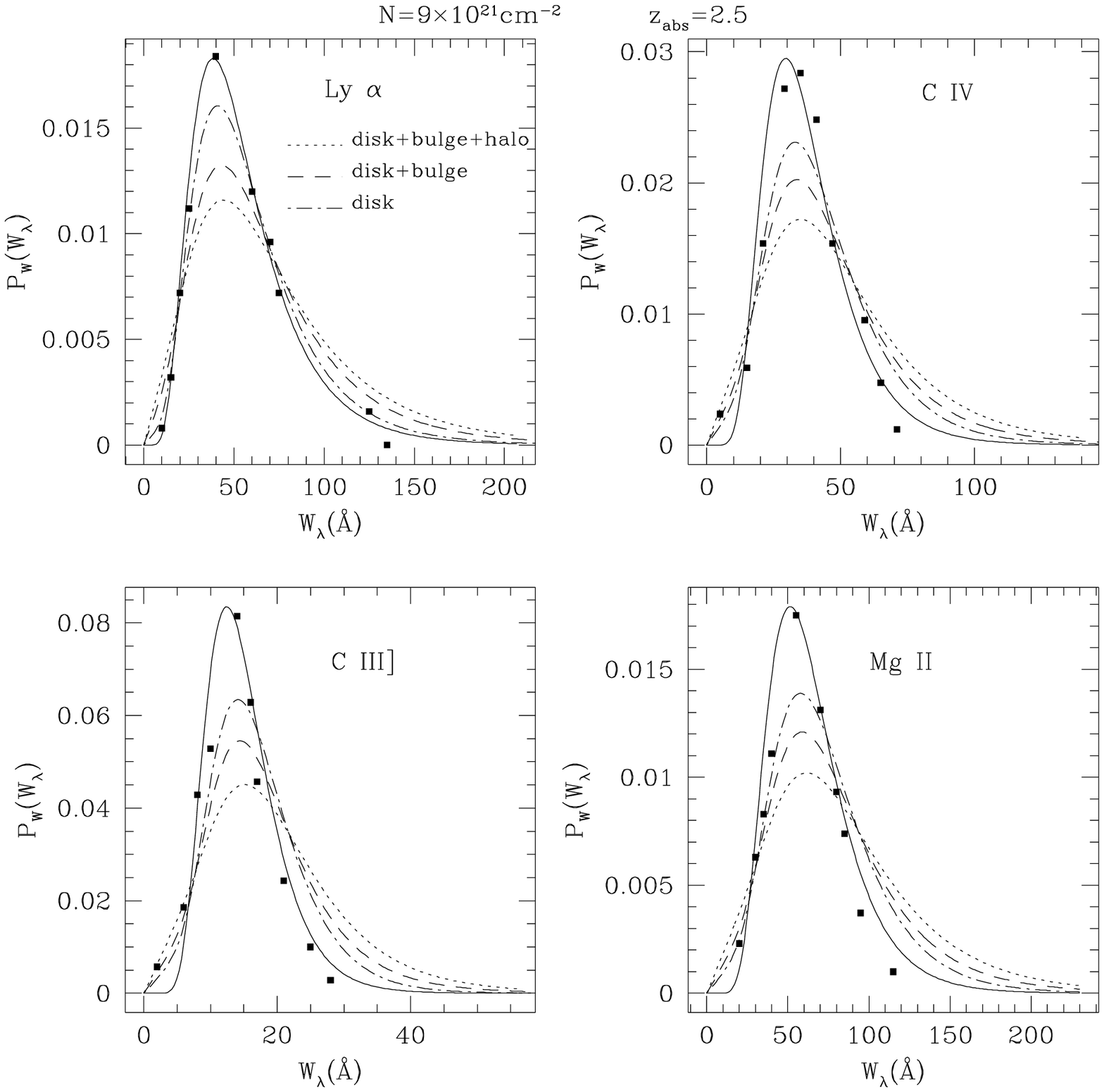}}
\caption{Same as Figure 4a, but with $N=9\times 10^{21}{\rm cm}^{-2}$.}
\label{fig:4b}
\end{figure}

\renewcommand{\thefigure}{\arabic{figmain}}
\setcounter{figmain}{5}
\begin{figure}[t]
\centerline{\epsfysize=5in\epsffile{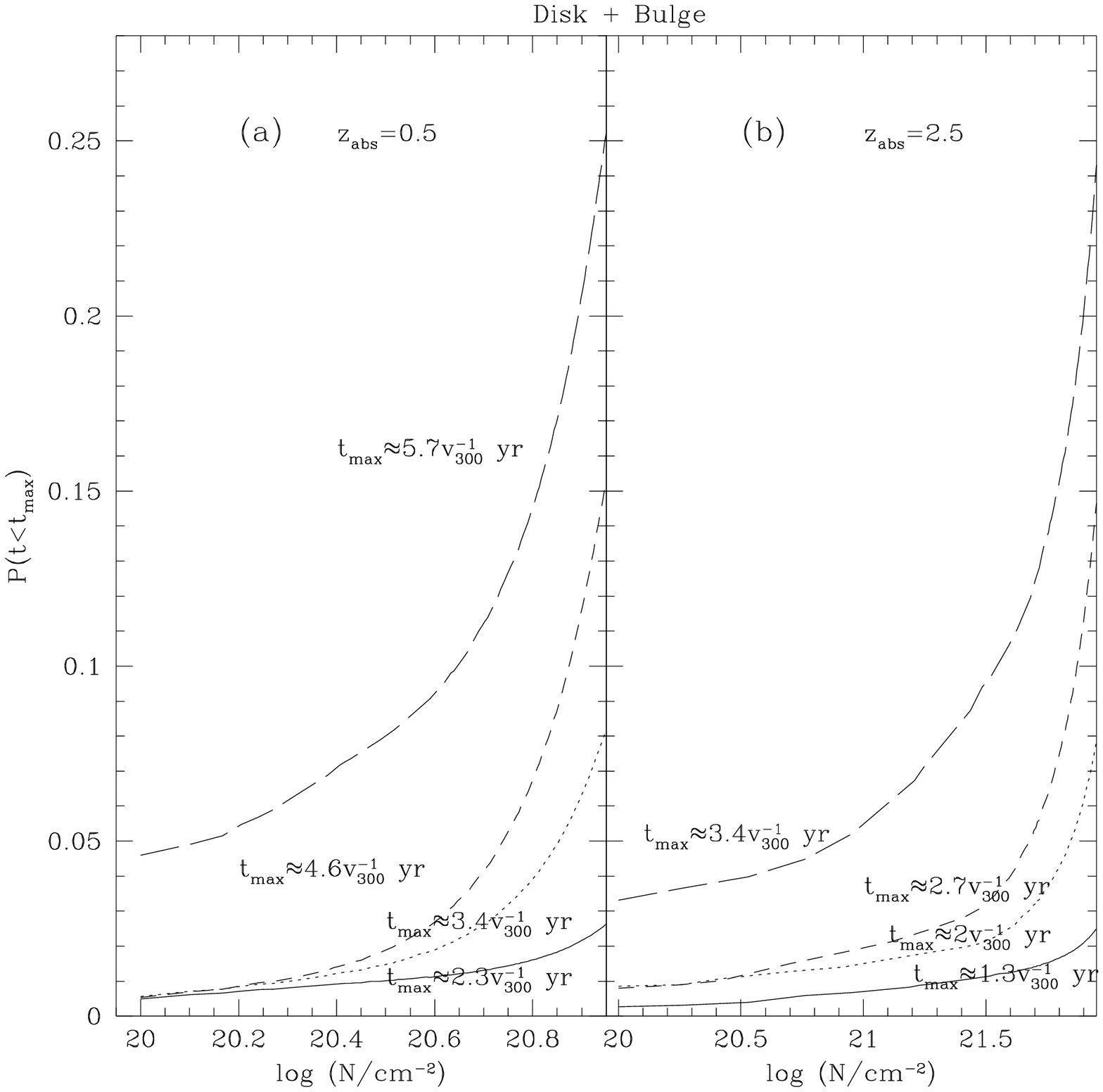}}
\caption{Probability for observing a magnification of 0.2 mag
with duration shorter than $t_{\rm max}$ as a function of the observed HI
column density.  Our model galaxy is located at a redshift $z_{\rm
abs}=0.5$ in panel (a) and $z_{\rm abs}=2.5$ in panel (b). In both panels,
$t_{\rm max}=0.08t_0$ (solid line), $t_{\rm max}=0.12t_0$ (dotted line),
$t_{\rm max} =0.16t_0$ (short dashed line), $t_{\rm max}=0.2t_0$
(long dashed line). Here, $t_0\approx 28v_{300}^{-1}\;{\rm yr}$~at $z_{\rm
abs}=0.5$, and $t_0\approx 17v_{300}^{-1}\;{\rm yr}$~at $z_{\rm abs}= 2.5$,
where $v_{300}$ is the transverse velocity of the microlenses relative to
the source in units of $300~{\rm km~s^{-1}}$.  These numbers were obtained
for the average stellar mass in a Scalo (1986) mass function.
The source redshift is
fixed at $z_{\rm s}=4$.}
\label{fig:5}
\end{figure}

\refstepcounter{figmain}
\begin{figure}[t]
\centerline{\epsfysize=5in\epsffile{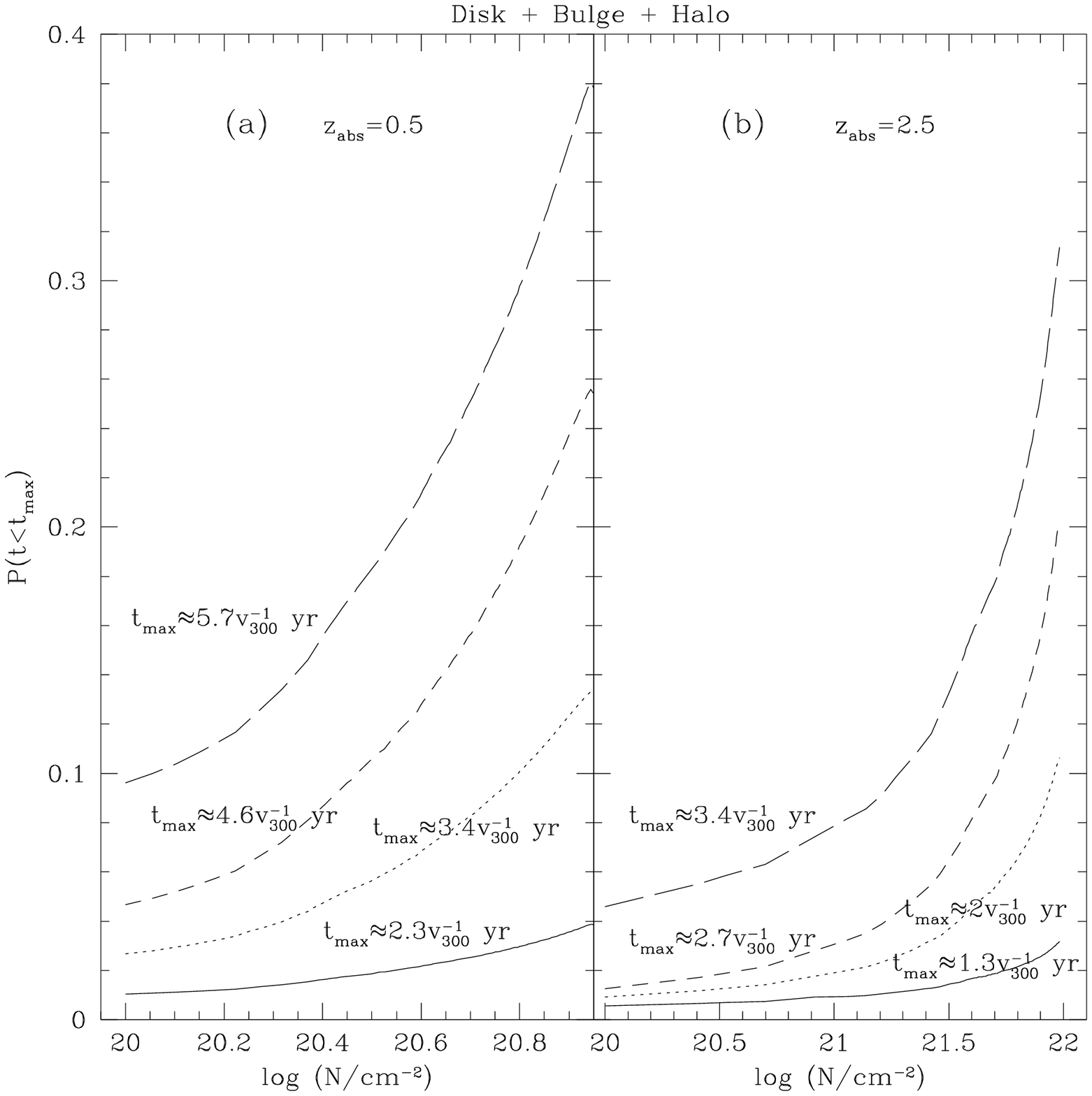}}
\caption{Same as Figure 5, but including
a halo made of MACHOs.}
\label{fig:6}
\end{figure}

\end{document}